# Improved Numerical Method for Calculation of 4-Body Transition Amplitudes

A L Harris[*]

Physics Department, Henderson State University, 1100 Henderson St, Arkadelphia, AR 71999, USA

**Abstract.**
In order to study 4-body atomic collisions such as excitation-ionization, transfer with target excitation, and double electron capture, the calculation of a nine-dimensional numerical integral is often required. This calculation can become computationally expensive, especially when calculating fully differential cross sections (FDCS), where the positions and momenta of all the particles are known. We have developed a new technique for calculating FDCS using fewer computing hours, but more memory. This new technique allows for much more efficient calculations and the use of many fewer resources.

## 1. Introduction

In recent years, there has been an increase in interest in 4-body atomic collisions. This is due in part to experimental advancements that allow for the measurement of fully differential cross sections (FDCS), which provide complete information about the positions and momenta of all particles in the collision. These more detailed experimental measurements have spurred theorists to develop state-of-the-art models for calculating the FDCS [1]-[15]. However, these calculations can present a significant computational challenge. This is due to the fact that as the number of particles in the collision grows, so does the number of degrees of freedom.

We have previously developed first order perturbative models for the 4-body processes of excitation-ionization, transfer-excitation, and double electron capture [1]-[3]. For each of these processes, our model requires the calculation of a 9-dimensional (9D) numerical integral, which is computationally expensive. To address the computational difficulty of our models, we have developed a new technique that allows us to perform the 4-body calculations with significantly fewer resources than in the past. We detail below how this technique can be applied to many first order perturbative 4-body models.

---

[*] harrisal@hsu.edu

Improved Numerical Method for Calculation of 4-Body Transition Amplitudes

## 2. Theory

To illustrate the technique, consider the case of electron-impact excitation-ionization (EI) of helium. In this process, a projectile electron collides with a helium atom. One of the target electrons is ionized and the other is left in an excited state of the $He^+$ ion.

In a first order perturbative model, the FDCS for EI is given by

$$FDCS = \mu_{pa}\mu_{ie} \frac{k_f k_e}{k_i} |T_{fi}|^2, \qquad (1)$$

where $\mu_{pa}$ is the reduced mass of the projectile and target atom, $\mu_{ie}$ is the reduced mass of the $He^+$ ion and the ejected electron, $\vec{k}_f$ is the momentum of the scattered projectile, $\vec{k}_e$ is the momentum of the ejected electron, $\vec{k}_i$ is the momentum of the incident projectile, and $T_{fi}$ is the transition matrix given by

$$T_{fi} = \langle \Psi_f | V | \Psi_i \rangle. \qquad (2)$$

The initial and final state wave functions are $\Psi_i$ and $\Psi_f$ respectively, and the perturbation is $V$. In the position representation, the transition matrix can be written as

$$T_{fi} = \int d\vec{r}_1 d\vec{r}_2 d\vec{r}_3 \Psi_f^*(\vec{r}_1,\vec{r}_2,\vec{r}_3) V \Psi_i(\vec{r}_1,\vec{r}_2,\vec{r}_3) \qquad (3)$$

where $\vec{r}_1, \vec{r}_2$, and $\vec{r}_3$ are the position vectors measured relative to the target nucleus for the projectile and two atomic electrons respectively (see figure 1).

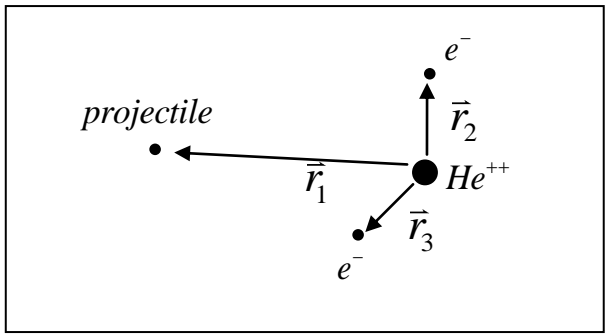

**Figure 1: Coordinate system for projectile-helium atom system.**

In the case of EI, we use our 4-Body Distorted Wave (4DW) model [1] to calculate the FDCS. In this model, the initial state wave function, $\Psi_i$, is given by

$$\Psi_i = \chi_i(\vec{r}_1)\xi_i(\vec{r}_2,\vec{r}_3). \tag{4}$$

where $\chi_i(\vec{r}_1)$ is a distorted wave for the incident projectile. This distorted wave is a solution of the Schrödinger equation for the distorting potential $U_i$, which is a spherically symmetric static potential calculated for the helium atom. The helium ground state wave function is given by $\xi_i(\vec{r}_2,\vec{r}_3)$, and can include both radial and angular correlation for the two bound electrons. The perturbation V is given by $V_i - U_i$, where

$$V_i = -\frac{2}{r_1} + \frac{1}{r_{12}} + \frac{1}{r_{13}} \tag{5}$$

is the initial state projectile-atom Coulomb interaction. The final state wave function $\Psi_f$, is given by

$$\Psi_f = \chi_f(\vec{r}_1)\chi_e(\vec{r}_2)\varphi_{nlm}(\vec{r}_3)C(\vec{r}_{12}), \tag{6}$$

where $\chi_f(\vec{r}_1)$ and $\chi_e(\vec{r}_2)$ are distorted waves for the scattered projectile and ionized electron respectively. They are solutions of the Schrödinger equation for the final state distorting potential $U_f$, which is given by a spherically symmetric approximation for the potential of the He$^+$ ion. It is calculated using a hydrogenic wave function for a nucleus of charge +2. The term $\varphi_{nlm}(\vec{r}_3)$ is the hydrogenic wave function for the He$^+$ ion in the $nlm$ state, and $C(\vec{r}_{12})$ is the Coulomb interaction between the two outgoing final state electrons. This Coulomb interaction is often referred to as post-collision interaction, or PCI. We note that for clarity we have labeled electron 2 as the ejected electron and electron 3 as the remaining bound electron of the He$^+$ ion. In our actual calculations, the final state wave function has been properly symmetrized.

With these wave functions, the transition matrix of equation (3) becomes

$$T_{fi} = \int d\vec{r}_1 d\vec{r}_2 d\vec{r}_3 \chi_f^*(\vec{r}_1)\chi_e^*(\vec{r}_2)\varphi_{nlm}^*(\vec{r}_3)C^*(\vec{r}_{12})(V_i - U_i)\chi_i(\vec{r}_1)\xi_i(\vec{r}_2,\vec{r}_3) \tag{7}$$

## 3. Numerical Technique

### 3.1 Original Method

Because all of the wave functions in equation (7) are calculated numerically, this 9D integral must be performed numerically. To do this, we use Gauss-Legendre quadrature integration for each of the 9 dimensions, as described in [16]. The Gauss-Legendre quadrature technique was chosen because the integrals for the FDCS are nearly symmetric about the momentum transfer direction. We are able to take advantage of this symmetry by orienting our z-axis along this direction, which allows us to use many fewer quadrature points for this coordinate than for the others. In a Monte Carlo approach, there would be no way to take advantage of this symmetry.

The calculation is performed using parallel processing with MPI, where the first of the 9 integration loops is split so that each quadrature point for the outermost integration loop is given to a different processor. Thus, each processor performs an 8D integral that corresponds computationally to 8 nested loops. After each processor completes its 8 integrations, the results are combined to give the final answer for the 9D integral. It is certainly possible to split the calculation among even more processors by placing the parallelization deeper into the nested integration loops, and in some of our codes this is done. However, for this demonstration, we consider only the case where the integration is split within the first loop.

The FDCS for EI are differential in incident projectile energy, scattering angle, ejected electron energy, and ejected electron angle. Thus, the 9D integral needs to be recalculated every time one of these physical parameters changes. A sample FDCS for EI is shown in figure 2. We note that to produce this curve, the FDCS was calculated at 25 different ejected electron angles between 0 and 360°.

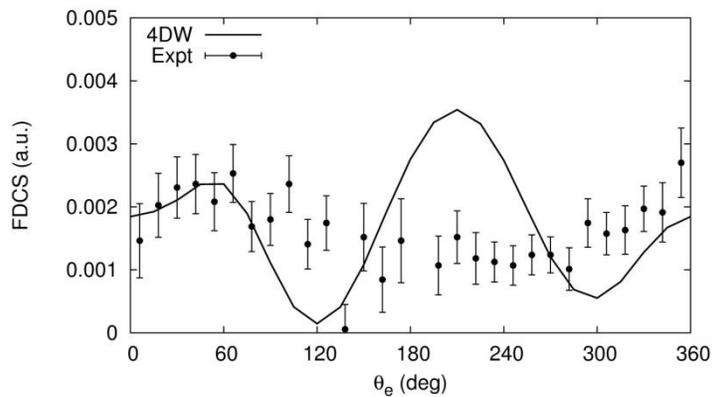

**Figure 2. FDCS for electron-impact excitation-ionization of helium to the 2p state as a function of ejected electron angle. Results are reproduced from [1] for an incident projectile energy of 500 eV, ejected electron energy of 10 eV, and scattering angle of 3.6 ° (solid line). Experimental data (solid circles) are from [17].**

In the past, performing FDCS calculations for many different physical parameters has not been feasible due to the vast amount of computational resources required for a single 4-body FDCS calculation. However, our new technique now allows us to perform the necessary 9D integral with significantly fewer resources.

**3.2 New Computational Method**

To more clearly describe this technique, consider the 9D transition matrix of equation (7). The dependence of the FDCS on the physical parameters is contained entirely in the incident projectile wave function $\chi_i(\vec{r}_1)$, the ejected electron wave function $\chi_e(\vec{r}_2)$, the scattered projectile wave function $\chi_f(\vec{r}_1)$, and the post-collision interaction $C(\vec{r}_{12})$. Because the $He^+$ wave

Improved Numerical Method for Calculation of 4-Body Transition Amplitudes

function $\varphi_{nlm}(\vec{r}_3)$, the helium ground state wave function $\xi_i(\vec{r}_2,\vec{r}_3)$, and the perturbation $(V_i - U_i)$ do not depend on these physical parameters, the integral over the $\vec{r}_3$ coordinate also does not depend on these physical parameters. This implies that the integral over $\vec{r}_3$ only needs to be performed once and stored as a function of the six coordinates defined by $\vec{r}_1$ and $\vec{r}_2$.

Equation (7) can then be rewritten as

$$T_{fi} = \iiint d\vec{r}_1 d\vec{r}_2 d\vec{r}_3 \chi_f^*(\vec{r}_1) \chi_e^*(\vec{r}_2) \chi_i(\vec{r}_1) C^*(\vec{r}_{12}) K^{EI}(\vec{r}_1,\vec{r}_2) \tag{8}$$

where $K^{EI}(\vec{r}_1,\vec{r}_2)$ is the integral over the $\vec{r}_3$ coordinate

$$K^{EI}(\vec{r}_1,\vec{r}_2) \equiv \int d\vec{r}_3 \varphi^*(\vec{r}_3)(V_i - U_i)\xi_i(\vec{r}_2,\vec{r}_3). \tag{9}$$

By storing $K^{EI}(\vec{r}_1,\vec{r}_2)$ as a function of $\vec{r}_1$ and $\vec{r}_2$, it can be used repeatedly in the calculation of $T_{fi}$, which is now simply the 6D integral of equation (8). The computational resources required to perform this 6D integral are quite minimal compared to the 9D integral of equation (7), as is demonstrated below.

## 4. Results

Figure 3 shows how the runtime for each processor of our original 4DW code scales as a function of the number of quadrature points $n$. All calculations were performed on the Stampede system at the University of Texas. For the analysis here, the number of quadrature points for each dimension is chosen to be the same. In reality, some dimensions need more points than others depending upon the amount of structure and oscillation in the integrand for that particular coordinate. The runtime for each processor is typically referred to as the wallclock time, and the total computing time can be found by multiplying the wallclock time by the number of processors used in the calculation. Because $n$ quadrature points are used for each of the 9 integrations, and the number of processors used is equal to the number of quadrature points in the outermost integration, the number of processors is simply equal to $n$.

Improved Numerical Method for Calculation of 4-Body Transition Amplitudes

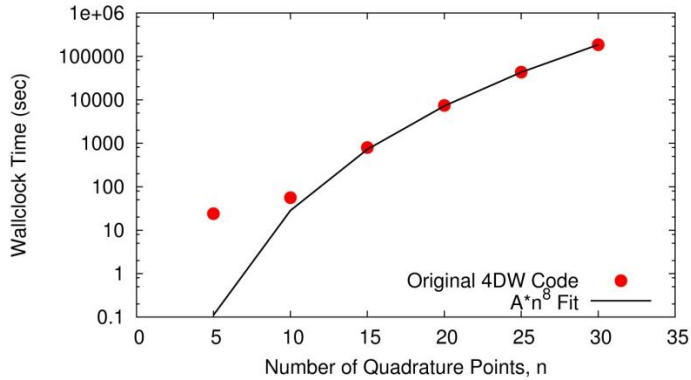

**Figure 3. (color online) Wallclock time of the original 4DW calculation as a function of the number of quadrature points (solid circles). The wallclock data is compared to the predicted scaling of the calculation (solid line).**

From figure 3, it can clearly be seen that the wallclock time of the calculation increases dramatically with the number of quadrature points used. For example, with $n = 10$, the wallclock time is 56 seconds, while for $n = 30$, the wallclock time is 52 hours. As mentioned in Section 3.1, each processor performs an 8D integral. Thus, it is expected that the wallclock time of the calculation should scale roughly as $An^8$, where $A$ is a fitting parameter that corresponds to the time for one calculation of the integrand. Figure 2 shows that asymptotically our code matches this fit, with the fitting parameter given by $A = 2.83 \times 10^{-7}$ s. The deviation of the data from the fit at small $n$ is due to the time associated with the setup of the terms inside the integrand. At low numbers of quadrature points, this setup time becomes comparable to the time for the integrand, and the real wallclock time of the calculation exceeds the $An^8$ scaling behavior.

To implement our new numerical technique, two separate calculations are required. The first calculation is to perform the integral over the $\vec{r}_3$ coordinate and store $K^{EI}(\vec{r}_1, \vec{r}_2)$ as a function of $\vec{r}_1$ and $\vec{r}_2$. This first step requires 9 nested loops, as in the original code, but this calculation only needs to be performed once. Figure 4 shows how the wallclock time of this storage part of the calculation scales with the number of quadrature points.

Improved Numerical Method for Calculation of 4-Body Transition Amplitudes

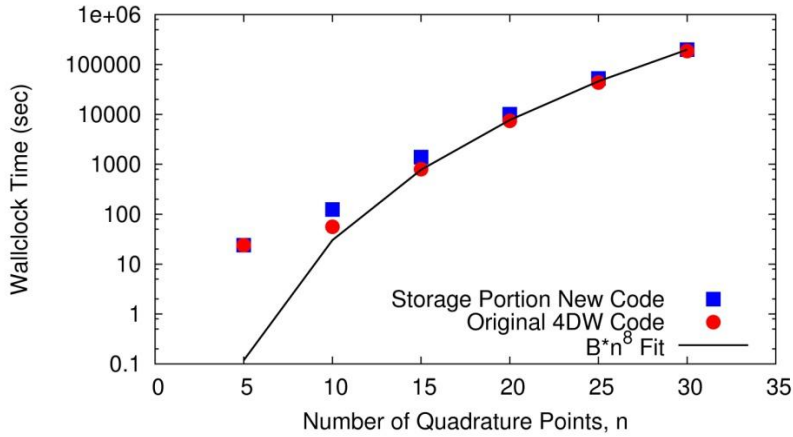

**Figure 4. (color online) Wallclock time of storage part of our numerical technique as a function of the number of quadrature points (solid squares). The data is compared to the wallclock time of the original code (solid circles) and the predicted scaling of the storage part of the calculation (solid line).**

Despite the fact that within the nested loops there is now a write statement, the wallclock time of this calculation is quite similar to that of the original 9D code. It is well known that writing to a file is an intrinsically slow process. However, because the write statement occurs after the innermost 3 loops have been completed, each processor only writes to the file $n^5$ times. Therefore, each processor performs $n^8$ calculations of the integrand and writes to the file $n^5$ times. This indicates that the scaling for the storage part of the technique should be approximately $B_1 n^8 + B_2 n^5$, where $B_1$ and $B_2$ are scaling parameters corresponding to the time for one calculation of the integrand and the time to complete one write statement respectively. Even at very small $n$, the scaling behavior is dominated by the $n^8$ term. Figure 4 confirms that the runtime of the storage part of the technique is quite similar to the original 9D code, and for large $n$, scales as $B_1 n^8$, with $B_1 = 3.02 \times 10^{-7}$ s.

The second calculation that is required for our new technique is to read the values of the stored integrand, and complete the remaining 6D integration of equation (9). This 6D integration is much less difficult computationally, and can be performed with a single processor in a matter of minutes. Because the read statement goes in the innermost loop of the 6 nested loops, the single processor must perform $n^6$ read operations and $n^6$ calculations of the integrand. It is therefore expected that the wallclock time for the 6D integration code should scale as $Cn^6$, where $C$ is a fitting parameter that corresponds to the time to read the value of $K^{EI}(\vec{r}_1, \vec{r}_2)$ once and the time to calculate the integrand once. A fit of the data results in $C = 2.86 \times 10^{-6}$ s. Figure 5 shows the scaling behavior of the wallclock time for the 6D integration part of the calculation.

Improved Numerical Method for Calculation of 4-Body Transition Amplitudes

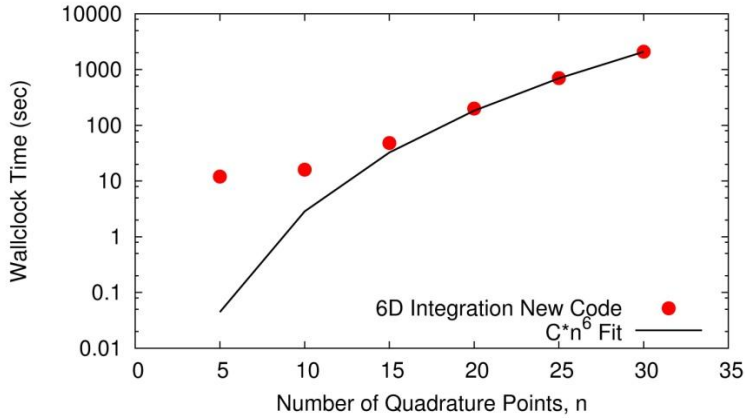

**Figure 5. (color online) Wallclock time of the 6D integration part of the calculation as a function of number of quadrature points (solid circles). The wallclock data is compared to the predicted scaling of this calculation (solid line).**

As in the case of the 9D codes, there is some setup time required for the various parts of the integrand, and at low numbers of quadrature points, this setup time becomes comparable to the integration time. Therefore, the wallclock time exceeds the $Cn^6$ scaling at low $n$, but approaches it at large $n$.

While our new technique dramatically saves on runtime, it does so at the expense of data storage. Because the 9D storage portion of the calculation writes one number to a file for all 6 coordinates of $\vec{r}_1$ and $\vec{r}_2$, the file size should scale as $Dn^6$, where $D$ is a fitting parameter that corresponds to the memory required to store one number in the file. A fit of the data results in $D = 100$ bytes. Figure 6 shows the file size as a function of the number of quadrature points, and it is indeed consistent with the predicted scaling.

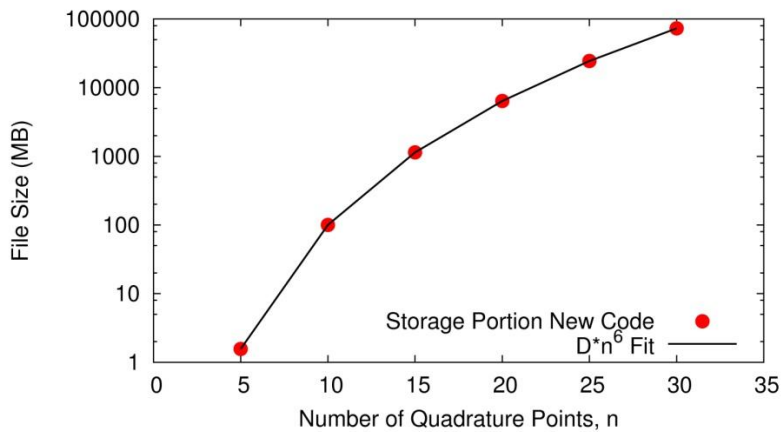

**Figure 6. (color online) File size as a function of the number of quadrature points (solid circles) compared to the predicted scaling of the file size (solid line).**

To get an idea of how much of an improvement this method provides, it is useful to look at the ratio of the wallclock time for the original calculation to the wallclock time for the new

Improved Numerical Method for Calculation of 4-Body Transition Amplitudes

methodology for a fixed value of $n$. This ratio can be thought of as the amount of "speed-up" achieved by the new calculation, and as more calculations are performed, the benefit of using the new method should increase. Figure 7 shows this ratio as a function of the number of calculations performed for a fixed value of $n = 30$ quadrature points.

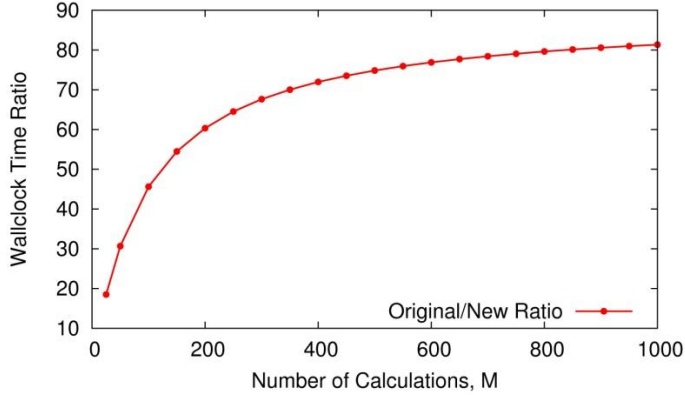

**Figure 7. (color online) Ratio of wallclock time of the original calculation to wallclock time of the new calculation as a function of the number of calculations performed. The number of quadrature points is fixed at $n = 30$.**

Based on the scaling trends of the original and new computational methodologies, one can see that this ratio should scale as the following:

$$ratio = \frac{M*(\text{original wallclock time})}{(\text{storage wall clock time}) + M*(\text{6D integration wallclock time})}$$

$$= \frac{M(An^8)}{B_1 n^8 + M(Cn^6)} \tag{10}$$

$$= \frac{1}{\left(\frac{B_1}{A}\right)\frac{1}{M} + \frac{C}{A}\frac{1}{n^2}}$$

where $M$ is the number of calculations performed. From the fits found in figures 3 through 6, $B_1/A = 1.07$ and $C/A = 10.1$. Then, the ratio of wallclock times becomes

$$ratio \approx \frac{1}{\frac{1}{M} + \frac{10}{n^2}}. \tag{11}$$

For a fixed number of quadrature points $n$ and large values of $M$, this ratio approaches $n^2/10$.

The results shown in Figure 7 were for a fixed value of $n = 30$ quadrature points, and thus the ratio levels off to a maximum value of about 90. In other words, after many calculations, the new method is 90 times faster than the original method. Even at the relatively small number of 25 calculations (the number used to create the graph of FDCS vs. ejected

Improved Numerical Method for Calculation of 4-Body Transition Amplitudes

electron angle in figure 2), the new method is 19 times faster than the original method. This results in a huge savings in number of computing hours required, and allows for the possibility of much broader studies of 4-body processes.

Excitation-Ionization is not the only process in which this method can be used to improve the efficiency of the calculations. The processes of transfer with target excitation, double electron capture, transfer-ionization, and double excitation can all be calculated using a method similar to the one described above. The only change required is the part of the 9D integral that is stored.

## 6. Conclusion

We have presented a new computational technique for the calculation of the transition amplitude for 4-body atomic collisions. This new technique results in a dramatic improvement in the amount of wallclock time required for such calculations, but increases the memory required for the calculation. For large numbers of calculations, the time required for the calculations decreased by a factor of 90, and at small numbers of calculations decreased by a factor of 19. The total memory required to use this method approaches 100 GB, which is quite manageable on today's supercomputing systems. This method can be applied to many different 4-body collision processes, and will result in a huge savings of computational hours.

**Acknowledgements**

This work used the Extreme Science and Engineering Discovery Environment (XSEDE) under grant number PHY-120027, which is supported by National Science Foundation grant number OCI-1053575. We would also like to acknowledge the support of the Arkansas Space Grant Consortium under grant number ASGC-HSU22087.

Improved Numerical Method for Calculation of 4-Body Transition Amplitudes